\documentclass{elsart}

\usepackage{natbib}
\usepackage{graphicx}
\usepackage{amssymb}
\usepackage{xspace}

\newcommand{\powersep}{{\ensuremath{\times}}}

\newcommand{\Msun}{{\ensuremath{\mathrm{M}_{\odot}}}\xspace}

\newcommand{\erg}{{\ensuremath{\mathrm{erg}}}\xspace}

\newcommand{\isofont}[1]{{\mathrm{#1}}}
\newcommand{\isomass}[1]{{\ensuremath{\isofont{^{#1}}}}}
\newcommand{\isocharge}[1]{{\ensuremath{\isofont{_{#1}}}}}
\newcommand{\isotope}[3]{{\ensuremath{\isocharge{#1}\isomass{#2}\isofont{#3}}}}

\newcommand{\I}[2]{{\isotope{}{#1}{#2}}}
\newcommand{\El}[1]{{\I{}{#1}}}

\newcommand{\Ep}[1]{{\ensuremath{10^{#1}}}}
\newcommand{\E}[1]{{\ensuremath{\powersep\Ep{#1}}}}

\newcommand{\Rag}{{\ensuremath{(\alpha,\gamma)}}\xspace}
\newcommand{\Ran}{{\ensuremath{(\alpha,\mathrm{n})}}\xspace}

\begin{document}

\begin{frontmatter}



\title{Massive Star Evolution: Nucleosynthesis and Nuclear Reaction
Rate Uncertainties}


\author[UCSC]{A. Heger},
\ead{alex@ucolick.org}
\author[UCSC]{S.E. Woosley},
\author[Basel]{T. Rauscher},
\author[LLNL]{R.D. Hoffman}, \\ \&
\author[UCSC]{M.M. Boyes}

\address[UCSC]{Department of Astronomy and Astropyhsics, University of
California, Santa Cruz, CA 95064} 
\address[Basel]{Department of Physics and Astronomy, University of
Basel, Basel, Switzerland} 
\address[LLNL]{Nuclear Theory and Modeling Group, Lawrence Livermore
National Laboratory, Livermore, CA 94550}

\begin{abstract}
We present a nucleosynthesis calculation of a 25\,\Msun star of solar
composition that includes all relevant isotopes up to polonium.  In
particular, all \emph{stable} isotopes and necessary nuclear reaction
rates are covered.  We follow the stellar evolution from hydrogen
burning till iron core collapse and simulate the explosion using a
``piston'' approach.  We discuss the influence of two key nuclear
reaction rates, \I{12}{C}\Rag and \I{22}{Ne}\Ran, on stellar evolution
and nucleosynthesis.  The former significantly influences the
resulting core sizes (iron, silicon, oxygen) and the overall
presupernova structure of the star.  It thus has significant
consequences for the supernova explosion itself and the compact
remnant formed.  The later rate considerably affects the
\textsl{s}-process in massive stars and we demonstrate the changes
that different currently suggested values for this rate cause.
\end{abstract}


\end{frontmatter}

\section{Introduction}
\label{intro}

Massive stars of more than 8\,\Msun are the main source of oxygen and
heavier elements in the universe.  Availability of (theoretical and
experimental) reaction rate data and computational resources make it
now possible to follow the complete nucleosynthesis in massive stars
from hydrogen burning till core collapse and through the supernova
explosion (\S\ref{nuc}).  However, significant uncertainties in
several key nuclear reaction rates still exist.  In \S\ref{ne22} we
discuss the influence of the \I{22}{Ne}\Ran rate on the
\textsl{s}-process in massive stars and in \S\ref{c12} we demonstrate
the influence of the \I{12}C\Rag rate on the presupernova structure.

\section{Complete nucleosynthesis study}
\label{nuc}
\begin{figure} 
\begin{center} 
\includegraphics*[width=\columnwidth, clip=false]{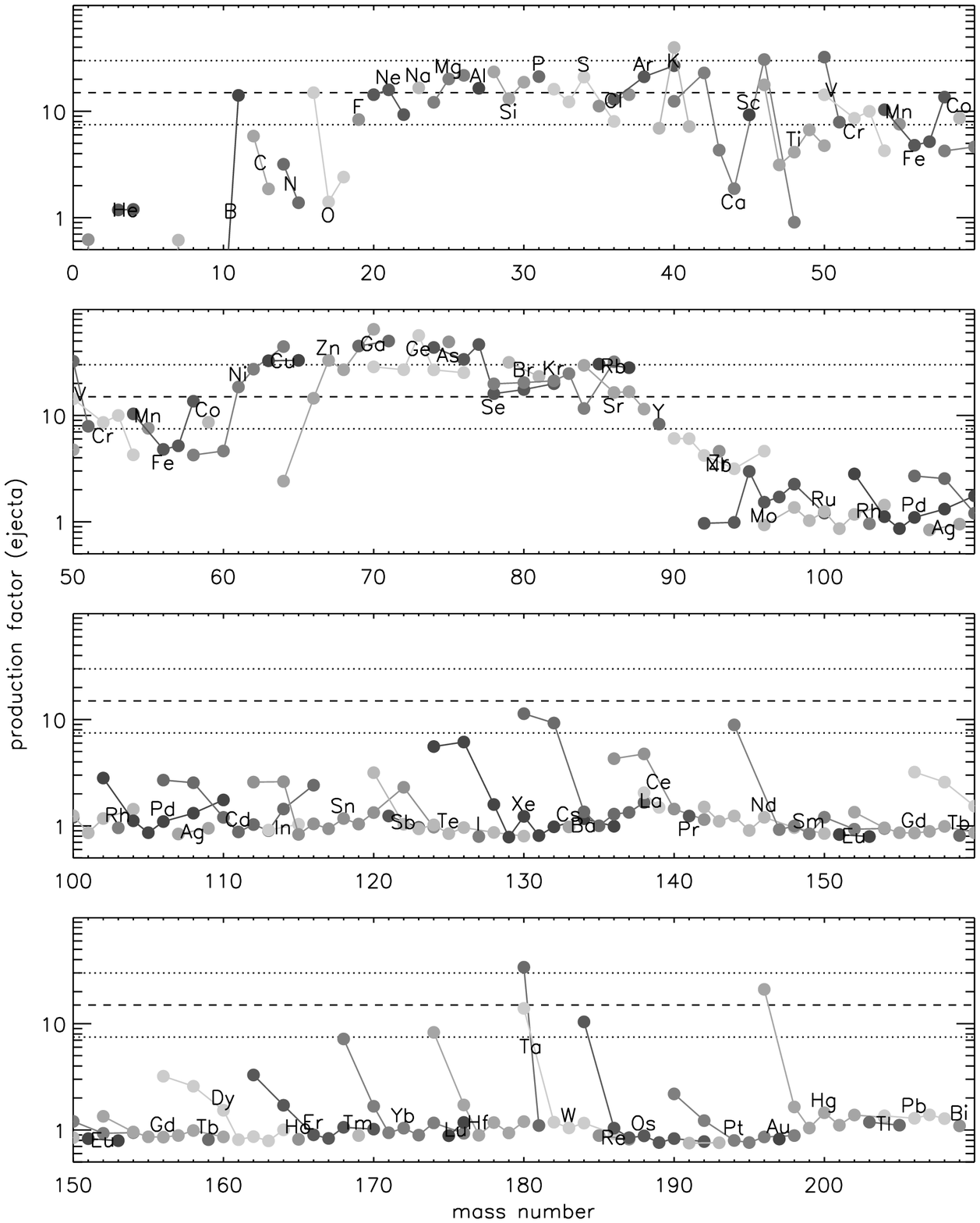} 
\end{center} 
\caption{Production factors of isotopes in ejecta, including wind,
relative to solar abundances for a 25\,\Msun star of solar
composition.  We use 1.2$\times$ the \I{12}C\Rag rate of \citet{Buc96}
\citep[cf.][]{kun01} and the low \I{22}{Ne}\Ran rate of \citet{kae94};
\cite[as used by][]{HWW00}.}
\label{fig:pf} 
\end{figure}

We present the first calculations to follow the complete
nucleosynthesis in massive stars from hydrogen ignition till onset of
iron core collapse and through the supernova explosion.
Figure~\ref{fig:pf} shows the average abundance of all ejecta,
including stellar wind mass loss, relative to their solar values
(production factor).  The dashed line indicates the production factor
for \I{16}O, the dominant ``metal'' produced by massive stars, and the
dotted lines indicate twice and half the production factor.  The
supernova explosion is simulated by a piston that resulted in
$\sim1.7\E{52}\,\erg$ kinetic energy of the ejecta. The mass cut is
then determined self-consistently from the hydrodynamical simulation.
Note that Fig.~\ref{fig:pf} does not include a possible
\textsl{r}-process contribution due to the neutrino wind from the
nascent neutron star.

In the 25\,\Msun star of Fig.~\ref{fig:pf}, most isotopes from oxygen
to the iron group are produced in about solar ratios relative to
oxygen.  The iron group itself is somewhat underproduced, due to
fall-back during the explosion.  Stars of $\sim15\,\Msun$ typically
contribute more here and Type Ia supernovae have added to the solar
abundance in the region.  The \textsl{s}-process isotopes above the
iron group till $\sim A=90$ are slightly overproduced.  Stars of lower
mass and/or lower metallicity produce less here.  Therefore these high
yields are required to produce the solar abundances over the lifetime
of the Galaxy.  Above $A\gtrsim100$ many \textsl{p}-isotopes are
produced by the $\gamma$-process in about solar abundances relative to
\I{16}O.  For more details please refer to \citet{rau01}.

\section{The \I{22}{Ne}\Ran rate}
\label{ne22}
\begin{figure} 
\begin{center} 
\includegraphics*[width=\columnwidth, clip=false]{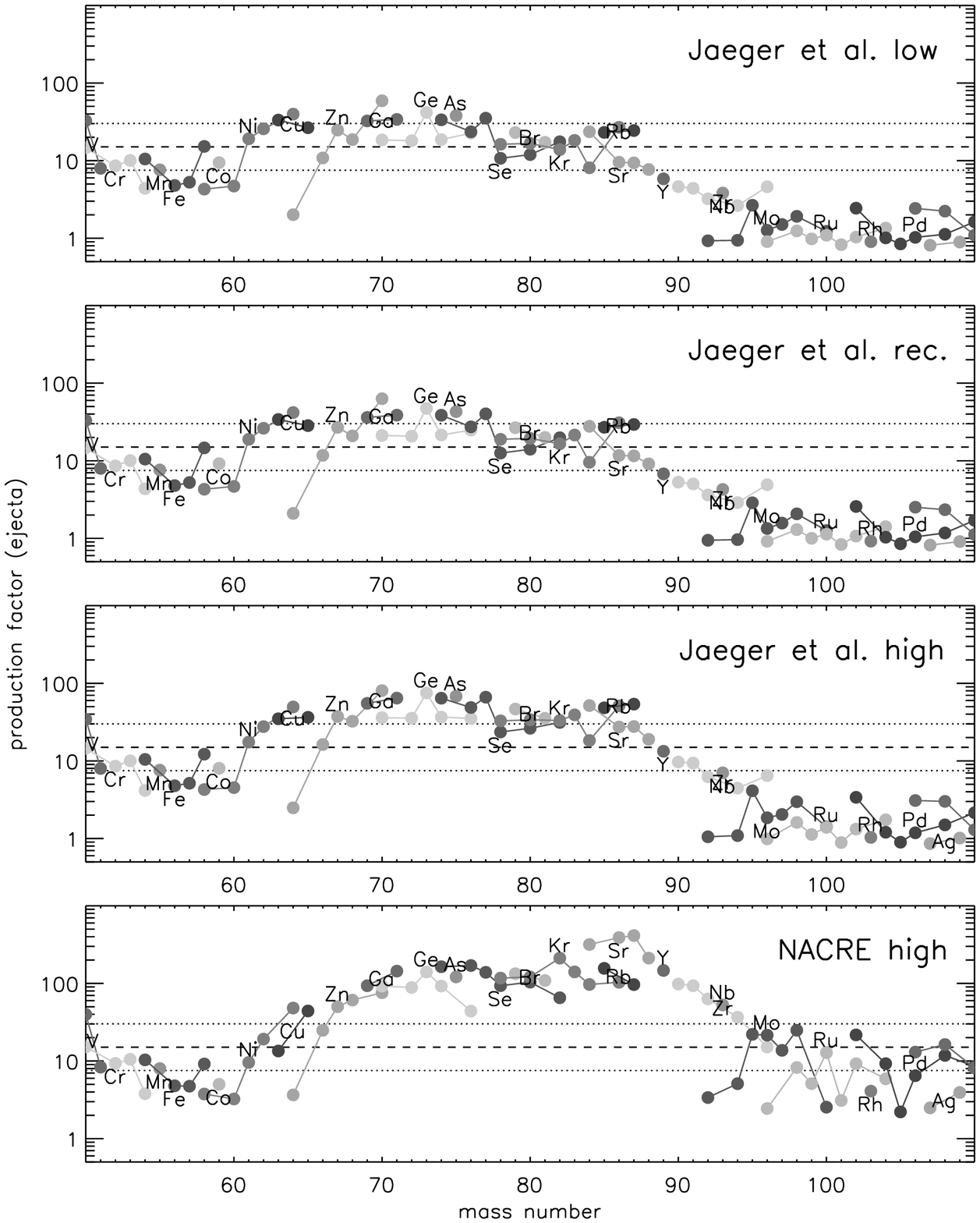} 
\end{center} 
\caption{Comparison of the production factors (see Fig.~\ref{fig:pf})
for different \I{22}{Ne}\Ran rates.  The first three panels give the
results for the lower limit, recommended value, and upper limit of
\citet{jae01}.  The bottom panel uses the reaction rate set by
\cite{NACRE} with their upper limit for the \I{22}{Ne}\Ran rate.}
\label{fig:ne} 
\end{figure}

The \I{22}{Ne}\Ran rate is the most important source of neutrons for
the \textsl{s}-process in massive stars in the region $60\lesssim A
\lesssim90$.  Figure~\ref{fig:pf} shows the result for the lower limit
given by \citet{kae94} \citep[as used by][]{HWW00}.  In
Fig.~\ref{fig:ne} we show the results for the same star, except that
we use the low, high, and recommended values by \citet{jae01}.  The
abundance ratio of the \textsl{s}-process does not change much by
this; the abundances only shifting to a slightly higher absolute
value.  The lower panel shows the result when using the high limit
given by \citet{NACRE} (and the rest of their rate set as well, which
makes only a minor difference as compared to \I{22}{Ne}\Ran).  The
large overproduction of the \textsl{s}-process here cannot be balanced
by galactochemical evolution.  Even so, we do not find significant
production of \textsl{p}-isotopes of \El{Mo} and \El{Ru}
\citep[cf.][]{cos00}.  For more details please refer to \citet{heg01}.

\section{The \I{12}C\Rag rate}
\label{c12}
\begin{figure} 
\begin{center} 
\includegraphics*[width=\columnwidth, clip=false]{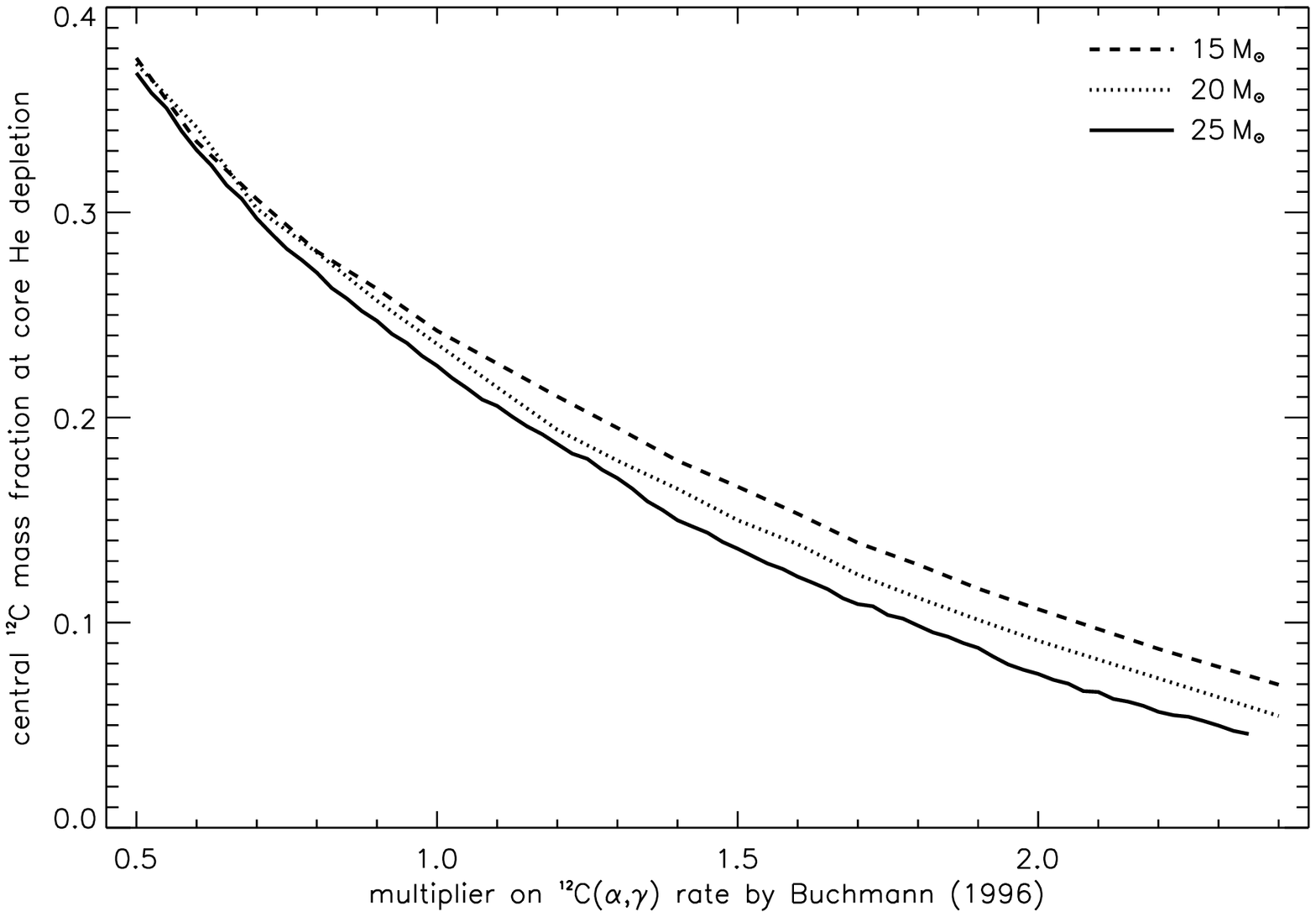} 
\end{center} 
\caption{Central carbon mass fraction after core helium exhaustion
as a function of a multiplier on the \I{12}C\Rag rate of \citet[, 2000 priv. com.]{Buc96}
for 15\,\Msun (\textsl{dashed line}), 20\,\Msun (\textsl{dotted
line}), and 25\,\Msun (\textsl{solid line}) stars.}
\label{fig:xcc} 
\end{figure}
\begin{figure} 
\begin{center} 
\includegraphics*[width=\columnwidth, clip=false]{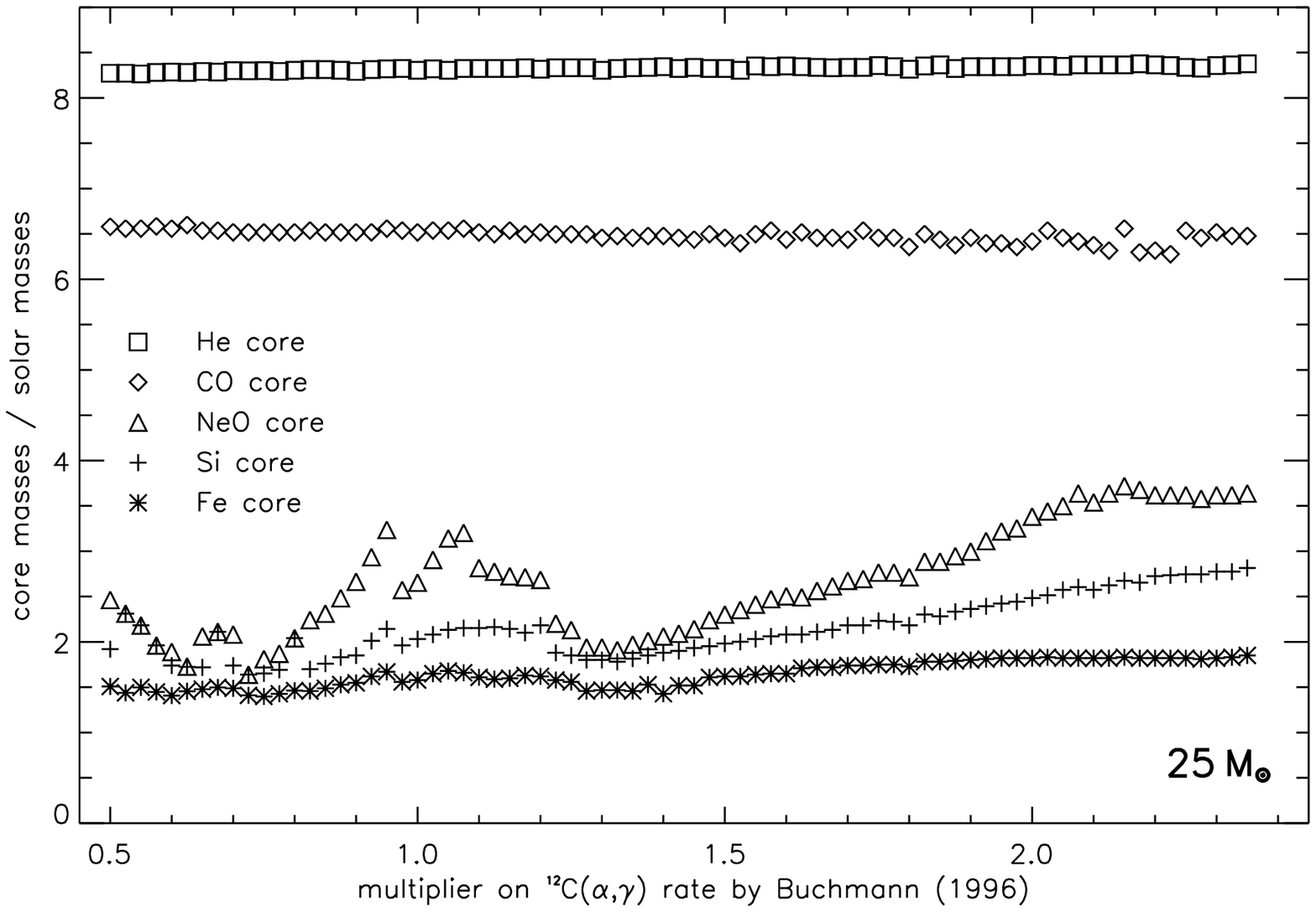} 
\end{center} 
\caption{Helium (\textsl{squares}), carbon-oxygen (\textsl{rotated
squares}), neon-oxygen (\textsl{triangles}), silicon
(\textsl{crosses}), and ``iron'' (\textsl{asterisks}) core masses
as a function of a multiplier on the \I{12}C\Rag rate of \citet{Buc96}}
\label{fig:core} 
\end{figure}

The uncertainty of the \I{12}C\Rag rate has significant influence on
the late time evolution of massive stars \citep[cf.][]{imb01}.  It
determines how much \I{12}C is left after core helium exhaustion
(Fig.~\ref{fig:xcc}).  Only for a sufficiently high value central
carbon burning proceeds convectively while otherwise it burns
radiatively.  Similarly, the extent and duration of the carbon shell
burning phases are affected.  They set the stage for the later burning
phases in that they produce carbon-free cores of different sizes, as a
non-monotonous function of the carbon abundance.  In
Fig.~\ref{fig:core} the iron core size varies by up to 30\,\%!
Therefore the \I{12}C\Rag rate determines whether a neutron star or a
black hole is formed.  For more details please refer to \citet{BHW01}.

\section{Conclusions \& outlook}
\label{con}

The current extent of uncertainties in key nuclear reaction rates
still has significant influence on stellar evolution and
nucleosynthesis.  Other major uncertainties comprise our understanding
of mixing processes, rotation, and magnetic fields in stars.  Some of
these uncertainties are currently under re-investigation.  The
combined effort of refined rate determinations and stellar modeling
allows both fields to profit from the synergy effects.

{\small
\textbf{Acknowledgments:}
This research was supported, in part, by the DOE (W-7405-ENG-48),
the National Science Foundation (AST 97-31569, INT-9726315), the
Alexander von Humboldt Foundation (FLF-1065004), and the Swiss
National Science Foundation (2000-061822.00).  T.R.\ acknowledges
support by a PROFIL professorship from the Swiss National Science
Foundation (2124-055832.98).
}

\end{document}